\documentclass[showpacs,pre,aps,twocolumn]{revtex4-1}

\pdfoutput=1
\usepackage{amsmath,amsfonts,amssymb,bm,graphicx,color}

\usepackage{hyperref}
\usepackage{color}

\renewcommand{\L}{\mathcal{L}}
\newcommand{\W}{\mathcal{W}}
\newcommand{\R}{\mathcal{R}}
\newcommand{\NJ}{N_{\rm J}}
\newcommand{\T}{\mathbb{T}}
\newcommand{\Iout}{\mathbb{I}_{\rm out}}
\newcommand{\A}{\mathbb{A}}
\newcommand{\B}{\mathcal{B}}
\newcommand{\G}{\mathcal{G}}
\newcommand{\V}{\mathcal{V}}
\renewcommand{\S}{\mathbb{S}}

\newcommand{\tL}{\tilde{\mathcal{L}}}
\newcommand{\tW}{\tilde{\mathcal{W}}}
\newcommand{\tR}{\tilde{\mathcal{R}}}

\newcommand{\er}[1]{Eq.~\eqref{#1}}
\newcommand{\ers}[2]{Eqs.~(\ref{#1}-\ref{#2})}
\newcommand{\era}[2]{Eqs.~(\ref{#1}) and (\ref{#2})}

\newcommand{\fs}{\langle - |}

\newcommand{\tobs}{t}

\begin{document}

\title{Classical stochastic dynamics and continuous matrix product states: gauge transformations, conditioned and driven processes, and equivalence of trajectory ensembles}

\author{Juan P. Garrahan}
\affiliation{School of Physics and Astronomy, University of Nottingham, Nottingham, NG7 2RD, United Kingdom}

\date{\today}

\begin{abstract}
Borrowing ideas from open quantum systems, we describe a formalism to encode ensembles of trajectories of classical stochastic dynamics in terms of continuous matrix product states (cMPSs).  We show how to define in this approach ``biased'' or ``conditioned'' ensembles where the probability of trajectories is biased from that of the natural dynamics by some condition on trajectory observables.  In particular, we show that the generalised Doob transform which 
maps a conditioned process to an equivalent ``auxiliary'' or ``driven'' process (one where the same conditioned set of trajectories is generated by a proper stochastic dynamics) is just a gauge transformation of the corresponding cMPS.  We also discuss how within this framework one can easily prove properties of the dynamics such as trajectory ensemble equivalence and fluctuation theorems.
\end{abstract}

\maketitle
\section{Introduction}

In this paper we describe a formalism that allows, given some classical stochastic Markovian dynamics \cite{Gardiner2004}, to encode the whole set of trajectories together with their probabilities of occurring---the {\em trajectory ensemble}---in terms of matrix product states (MPSs) \cite{[For a review see ]Verstraete2008}.  As we explain below, this approach enables to derive in a compact and concise way many (known) properties of trajectory ensembles associated to dynamical large-deviations  \cite{Jack2010,Garrahan2010,Chetrite2013,Chetrite2015} and non-equilibrium fluctuations  \cite{Chetrite2011,[For a review see ]Seifert2012}. 

As originally conceived \cite{Verstraete2008}, MPSs correspond to a variational class of quantum states much used to describe properties of one-dimensional quantum systems with low entanglement, such as ground states of quantum spin chains away from a quantum phase transition point.  In this context, the physical space is the one-dimensional space where the system lives, for example a one dimensional lattice of sites containing the spins. An MPS is then constructed as a superposition of basis states on the physical space (say classical or Fock states) where the amplitudes are given by the matrix elements of products of matrices that act on an auxiliary or ``bond space''. 
The scope of the variational class is determined by the dimension of the bond space, and when this auxiliary dimension is large enough quantum states can be described accurately via MPSs.  In this same spirit, the stationary distributions of certain one-dimensional classical stochastic processes can be described exactly in terms of MPSs with infinite bond dimension, most notably in the case of the simple exclusion process \cite{Derrida1993,Blythe2007}.  

More recently, MPSs have been extended \cite{Verstraete2010,Haegeman2013} to account for one-dimensional systems which live in the continuum, that is for one-dimensional quantum fields.  This generalisation, termed continuous matrix product states (cMPSs), can be thought of as a limit of MPSs where the lattice spacing of their physical space tends to zero; see \cite{Verstraete2010,Haegeman2013} for details.  Interestingly, cMPSs have an alternative interpretation \cite{Verstraete2010,Osborne2010} in terms of the input-output formalism \cite{Gardiner2004b} for open quantum systems (i.e., quantum systems interacting with an environment) undergoing continuous quantum Markovian dynamics.
In this interpretation the bond space is the physical space where the system lives, and the continuous one-dimensional space is the ``output'', a time record of the interaction between system and environment.  It is possible in this way to represent the whole ensemble of quantum trajectories of the open system as a cMPS.  This MPS approach can be then extended for the application of large-deviation methods to describe and characterise open quantum dynamics \cite{Lesanovsky2013,Kiukas2015}.

Here we adapt these ideas to classical stochastic dynamics
(for concreteness we focus on dynamics described by continuous time Markov chains).  We show that one can encode ensembles of classical trajectories in terms of cMPSs which, in contrast to the quantum case, are real rather than complex vectors.  A single cMPS then encodes all possible trajectories of the dynamics and their probabilities of occurring.  These cMPSs can be generalised to account for {\em biased} or {\em conditioned} trajectory ensembles \cite{Lecomte2007,Garrahan2007,*Garrahan2009,Chetrite2015}, i.e., sets of trajectories whose probabilities are altered from those of the natural dynamics according to some condition on trajectory observables.  These conditioned ensembles are of interest \cite{Lebowitz1999,Lecomte2007,Garrahan2007,Evans2004,Merolle2005,Maes2008,Baiesi2009,Hedges2009,Kurchan2009,Nemoto2011,Giardina2011,Budini2011,Speck2012,Speck2012b,Catana2012,Bodineau2012,Flindt2013,Weber2013,Espigares2013,Mey2014,Vaikuntanathan2014,Weber2015,Gingrich2016,Jack2015,Ueda2015,DeBacco2015,Szavits2015,Verley2016,Bonanca2016,Nemoto2016,Speck2016,Jack2016} as they characterise information about fluctuations of the dynamics away from typical behaviour. For certain forms of conditioning, biased trajectory ensembles can be  generated by some (tilted) operator, but in general this operator is non-stochastic (in the sense it does not conserve probability).  An important question is how to relate a conditioned trajectory ensembles to one generated by some other properly stochastic dynamics, termed the ``auxiliary'' or ``driven'' ensemble \cite{Jack2010,Garrahan2010,Chetrite2015}.  The mapping between conditioned and driven ensembles is the so-called {\em generalised Doob transform} \cite{Chetrite2015}. 

We will show how all these issues can be discussed in the context of cMPSs.  In particular, the key property we will exploit is that of the {\em gauge symmetry} of cMPSs \cite{Haegeman2013}, that is, the equivalence between different cMPSs vectors under a general class of time-dependent transformations.  We introduce the cMPS formalism for classical dynamics in Sec.\ II, and extend it to conditioned trajectory ensembles in Sec.\ III.  In Sec.\ IV we discuss cMPS gauge transformations and show how they can be used to obtain the generalised Doob transform \cite{Chetrite2015} between conditioned and driven ensembles.  Section V discusses the equivalence of trajectory ensembles \cite{Chetrite2013} in this context, and in Sec.\ VI we describe how one can obtain (integral) fluctuation relations \cite{Seifert2012} from the cMPS gauge symmetry.   In Sec.\ VII we give our conclusions.

\section{Ensembles of stochastic trajectories and matrix product states}

Consider a classical stochastic system evolving as a continuous time Markov chain.  The Master Equation (ME) for the probability reads~\cite{Gardiner2004}
\begin{align}
\partial_{t} P(C,t) = \sum_{C' \neq C} W({C' \to C}) P(C',t) 
\nonumber \\
- R(C) P(C,t) , 
\label{ME1}
\end{align}
where $P(C,t)$ indicates the probability of the system being in configuration $C$ at time $t$, 
$W({C' \to C})$ is the transition rate from $C'$ to $C$, and $R({C}) = \sum_{C' \neq C} W({C \to C'})$ the escape rate from $C$. The ME can be written in operator form,  
\begin{equation}
\partial_{t} |P(t) \rangle = \L |P(t) \rangle ,
\label{ME2}
\end{equation}
with probability vector $|P(t)\rangle$ 
\begin{equation}
|P(t)\rangle = \sum_{C} P(C,t) | C \rangle
\label{P} ,
\end{equation}
where $\{ | C \rangle \}$ is an orthonormal configuration basis, $\langle C | C' \rangle = \delta_{C,C'}$.  The Master operator $\L$ is defined as,
\begin{align}
\L &= \sum_{C,C' \neq C} W({C \to C'}) |C' \rangle \langle C|
- \sum_{C} R({C}) |C \rangle \langle C| 
\nonumber
\\
&= \sum_{\mu=1}^{N_{\rm J}} \W_{\mu} - \R ,
\label{L}
\end{align}
where by $\mu$ we label all possible transitions $C \to C'$, so that $\{ \W_{\mu} \}$ stand for the {\em jump operators}, $\{ \W_{C \to C'} = W({C \to C'}) |C' \rangle \langle C| \, , \,\, \forall \, (C \neq C') \}$.  $\R$ is the {\em escape rate operator}, $\R = \sum_{C} R({C}) |C \rangle \langle C|$.  If the dimension of configuration space is $D$ (e.g.\ $D=2^{N}$ in a system of $N$ classical Ising spins), then the number of jump operators is at most $N_{\rm J}=D(D-1)$.   

The dynamics described by the ME is realised by stochastic trajectories.  Each trajectory is a particular realisation of the noise that gives rise to a time record of configurations and of waiting times for jumps between them, observed up to a time $\tobs$.  We denote such a trajectory that starts in configuration $C_{0}$ by $\omega_t = ( C_{0} \to C_{t_1} \to \ldots \to C_{t_K} )$.  Here $t_{i}$ is the time when the transition from $C_{t_{i-1}}$ to $C_{t_{i}}$ occurs, so that the waiting time for the $i$-th jump is $t_{i}-t_{i-1}$.  The trajectory $\omega_t$ has a total of $K$ configuration changes (and $t_K \leq t$, i.e., between $t_K$ and $t$ no jump occurred).  Given the initial configuration, the trajectory is thus fully determined by the sequence of jumps $[(\mu_1,t_{1}), (\mu_2,t_{2}), \cdots]$, where the pair $(\mu_{i},t_{i})$ indicates a jump of kind $\mu_i$ occurring at time $t_{i}$. 

The probability of a trajectory $\omega_{t}$ can be written as a ``matrix element'',
\begin{equation}
{\rm Prob}(\omega_{t}) = \fs \V(\omega_{t}) | C_{0} \rangle ,
\label{Pomega}
\end{equation}
where $\V(\omega_{t})$ is a product of operators acting on the system, 
\begin{align}
\V(\omega_{t}) = e^{- (t-t_K) \R}
\,
\W_{\mu_K} e^{- (t_K-t_{K-1}) \R} 
\cdots
\nonumber \\
\cdots 
\W_{\mu_{2}} e^{- (t_{2}-t_{1}) \R}
\,
\W_{\mu_{1}} e^{- t_{1} \R} ,
\label{VV}
\end{align}
and $\fs$ is the ``flat'' or ``trace'' state, $\fs = \sum_{C} \langle C|$.  Note that $\fs P(t)\rangle = 1$ 
at all times due to normalisation of the probability.  In fact, $\fs$ is the left eigenstate of $\L$ with eigenvalue zero, $\fs \L = 0$, which is the statement of probability conservation. 

We now adapt the procedure of Refs.\ \cite{Verstraete2010,Osborne2010,Haegeman2013} to our problem of classical stochastic dynamics. 
To each trajectory we can associate a vector in an auxiliary or {\em output} space. 
This is an abstract space that will allow to keep a record of the specific sequence of events in a trajectory.  That is, given a trajectory $\omega_t$ due to the sequence of jumps $[(\mu_1,t_{1}), (\mu_2,t_{2}), \cdots, (\mu_K,t_K)]$, we can define a state $| \omega_{t} \rangle$ in the output,
\begin{equation}
| \omega_{t} \rangle = a_{\mu_K}^{\dagger}(t_K) \cdots a_{\mu_{2}}^{\dagger}(t_{2}) \,
a_{\mu_{1}}^{\dagger}(t_{1}) \,
| \Omega \rangle ,
\label{om}
\end{equation}
where we have introduced a one-dimensional system of bosons of $\NJ$ different kinds, with field operators $a_\mu(t)$ ($\mu = 1, \ldots , \NJ$) obeying commutation relations $[ a_\mu(t) ,  a_{\mu'}^\dagger(t') ] = \delta_{\mu \mu'} \delta(t-t')$.  
The state $| \Omega \rangle$ is the vacuum of these bosonic fields, $a_\mu(t) | \Omega \rangle = 0 \; \forall \mu$.  With these definitions, a trajectory is encoded through \er{om} by creating a ``particle'' in the output state that indicates both the kind of jump and its time: if a jump of kind $\mu_{i}$ occurs at time $t_{i}$, the field operator $a_{\mu_{i}}^\dagger(t_{i})$ creates a corresponding particle of type $\mu_{i}$ in the output space at ``position'' $t_{i}$.  Note that these are not meant to represent real particles but to associate a vector to each trajectory of the dynamics. (As we are considering continuous time dynamics such vectors are mathematically analogous to states of a bosonic field theory, see \cite{Verstraete2010,Osborne2010,Haegeman2013}).

If we combine the system and output state we can encode the set of trajectories in a single vector, $| \Psi_t \rangle = \sum_{\omega_t} \V(\omega_t) | C_0 \rangle \otimes | \omega_t \rangle$, which written explicitly becomes (cf.\ \cite{Verstraete2010,Osborne2010,Haegeman2013}), 
\begin{widetext}
\begin{align}
| \Psi_t \rangle 
=
\sum_{K=0}^\infty 
\,
\sum_{\mu_1= 0}^{\NJ} \cdots \sum_{\mu_K= 0}^{\NJ} 
\,
\int_0^t dt_1 
\int_{t_1}^t dt_2
\cdots
\int_{t_{K-1}}^t dt_K
\, 
e^{- (t-t_K) \R}
\,
\W_{\mu_K} e^{- (t_K-t_{K-1}) \R} 
\cdots
\W_{\mu_{1}} e^{- t_{1} \R} | C_0 \rangle 
\nonumber \\
\otimes
\, 
a_{\mu_K}^{\dagger}(t_K) \cdots a_{\mu_{2}}^{\dagger}(t_{2}) \,
a_{\mu_{1}}^{\dagger}(t_{1}) \,
| \Omega \rangle ,
\label{MPS1}
\end{align}
\end{widetext}
where the sum over all trajectories $\omega_t$ is a sum over all possible number of jumps $K$, of all kinds $\mu$, occurring at all possible times between $0$ and $t$.  The state $| \Psi_t \rangle$ is a (continuous) matrix product state (cMPS).  Equation \eqref{MPS1} shows that {\em the whole ensemble of trajectories of the dynamics can be encoded in a cMPS}.  
Furthermore, the cMPS $| \Psi_t \rangle$ can be written compactly as a path ordered exponential (cf.\ \cite{Verstraete2010,Osborne2010,Haegeman2013}),
\begin{equation}
| \Psi_t \rangle =
\T 
e^{ 
\int_0^t dt' 
\left[ 
\sum_{\mu=1}^{\NJ} \W_\mu \otimes a_\mu^\dagger(t') 
- \R \otimes \Iout 
\right]
}
| C_0 \rangle \otimes | \Omega \rangle ,
\label{MPS2}
\end{equation}
where $\Iout$ is the identity in the output space and $\T$ is the time ordering operator.  
Note the following:

\smallskip $\bullet$ 
The state $| \Psi_t \rangle$ is a real and positive vector.  Its coefficients are probabilities (rather than complex amplitudes as in the quantum case \cite{Verstraete2010,Osborne2010,Haegeman2013}). For example, the probability of a trajectory $\omega_t$ is obtained as 
\begin{equation}
{\rm Prob}(\omega_t) = \langle - | \otimes \langle \omega_t | \Psi_t \rangle .
\label{N1}
\end{equation}

\smallskip  $\bullet$ 
In analogy with the trace state of the system, we can define a trace state for the output, $\langle -_{\rm out} | = \sum_{\omega_t} \langle \omega_t |$.  If we trace out the output in 
$| \Psi_t \rangle$ we recover the average evolution of the system, as described by the ME of \er{ME2},
\begin{equation}
\langle -_{\rm out} | \Psi_t \rangle = e^{t \L} | C_0 \rangle .
\label{N2}
\end{equation}

\smallskip  $\bullet$ 
Tracing over both system and output we can see that $| \Psi_t \rangle$ is normalised to unity,
\begin{equation}
\langle -_{\rm tot} | \Psi_t \rangle = 1 ,
\label{N3}
\end{equation}
where $\langle -_{\rm tot} | = \langle - | \otimes \langle -_{\rm out} |$.  
The above result is due to probability conservation.  Note that since we are describing a classical system the space to which $| \Psi_t \rangle $ belongs is an L1-space (where the norm of a vector $|x\rangle$ is the 1-norm, $\Vert x \Vert_1 = \sum_k |x_k|$, rather than the 2-norm of a Hilbert space). 

\smallskip  $\bullet$ 
While the definitions above are for a specific initial state $| C_0 \rangle$, any initial distribution of configurations can of course be considered by replacing $| C_0 \rangle$ by a probability vector $| P_0 \rangle$. 

\smallskip  $\bullet$ 
In the discussion leading to Eqs.\ \eqref{MPS1}-\eqref{MPS2} we have considered dynamics generated by a time-independent Master operator $\L$.  Equations \eqref{MPS1}-\eqref{MPS2} are equally valid if the operators $\W_\mu$ of $\R$ are time-dependent.  We will elaborate on this point below.

\subsection{Example}
\label{Ex}

We illustrate the ideas presented so far with an elementary example which is exactly solvable.   We consider a particle that hops between nearest neighbouring sites of a one-dimensional lattice with periodic boundary conditions.  The set of configurations is given by the possible positions of the particle, so that  $\{ | C \rangle \} = \{ | x \rangle :  \; x=1,\ldots,L \}$.  Lets consider the case where hopping is fully asymmetric and all hopping rates are the same.  We can then label the jump operators $\W_{\mu}$ by the sites $x$, so that 
\begin{equation}
\W_{x} = \gamma |x+1\rangle \langle x|
\label{Ex1}
\end{equation}
is the operator corresponding to a jump from site $x$ to site $x+1$, where we identify $|L+1\rangle$ with $|1\rangle$ to implement the periodic boundary conditions.   The corresponding escape rate operator is proportional to the identity,
\begin{equation}
\R = \gamma \sum_{x=1}^{L} |x \rangle \langle x| .
\label{Ex2}
\end{equation}
The generator $\L$ reads,
\begin{equation}
\L = \gamma \sum_{x=1}^{L} |x+1\rangle \langle x| - \gamma \sum_{x=1}^{L} |x \rangle \langle x| .
\label{Ex3}
\end{equation}
	
The cMPS for this elementary dynamics takes a simple form.  Since $\R$ is proportional to the identity, all the waiting time factors can be commuted through all the jump operators in \er{MPS1}.  Furthermore, if the particle is at site $x$ the only possible jump is to site $x+1$, so that if the starting position is $x_0$ then the only non-vanishing sequence of $K$ jump operators is the product $\W_{x_{0}+K} \W_{x_{0}+K-1} \cdots \W_{x_{0}+1} \W_{x_{0}}$.  This means that given the initial position, all that matters to determine uniquely a trajectory are the times of the jumps.  In this sense the jump label $\mu$ in the creation operators $a^{\dagger}_{\mu}$, cf.~\er{om}, is redundant since which jump takes place is implicit in the order of the jump.  The Dyson series representation of the cMPS, \er{MPS1}, simplifies to,
\begin{align}
| \Psi_t \rangle 
=
e^{- t \gamma}
\sum_{K=0}^\infty 
\,
\int_0^t dt_1 
\cdots
\int_{t_{K-1}}^t dt_K \, \W^{K} | x_0 \rangle
\nonumber \\
 \otimes \, a^{\dagger}(t_K) \cdots a^{\dagger}(t_{1}) \,
| \Omega \rangle ,
\end{align}
where we have defined
\[
\W = \sum_{x} \W_{x} ,
\] 
since 
\[
\W^{K} |x_{0}\rangle = \W_{x_{0}+K} \cdots \W_{x_{0}}|x_{0}\rangle .
\]
In turn the equivalent path ordered exponential, \er{MPS2}, reads, 
\begin{equation}
| \Psi_t \rangle =
e^{-\gamma t} \;
\T 
e^{\int_0^t dt' \W \otimes a^\dagger(t')}
| x_0 \rangle \otimes | \Omega \rangle .
\end{equation}

We can interrogate the cMPS to extract for example the probability of a trajectory $\omega_{t}$, cf.~\er{N1}.  Since $\langle - | \W = \gamma \langle - |$ we obtain that 
\begin{equation}
{\rm Prob}(\omega_t) = e^{-\gamma t} \gamma^{K} ,
\label{ExPo}
\end{equation}
irrespective of when the jumps occur, where $K$ is the number of jumps in the trajectory.  This is of course the result expected for a Poisson process, which is what the dynamics of the fully asymmetric jumping particle is.

\section{Trajectory observables and conditioned ensembles}

In dynamics we are often interested in the behaviour of trajectory observables \cite{Lecomte2007,Garrahan2007}, i.e., quantities which depend on the whole dynamical trajectory and not just on the system at any one particular time.  Examples include the dynamical activity \cite{Lecomte2007,Garrahan2007,Baiesi2009},
\begin{equation}
K(\omega_t) = \sum_\mu n_\mu(\omega_t) ,
\end{equation}
which corresponds to the total number of configuration changes in a trajectory $\omega_t$, where $n_\mu(\omega_t)$ counts the number of jumps of kind $\mu$; or an integrated current \cite{Seifert2012},
\begin{equation}
J(\omega_t) = \sum_{(C,C')} [n_{C \to C'}(\omega_t) - n_{C' \to C}(\omega_t)],
\end{equation}
which is an accumulated difference in forward/backward transitions between certain pairs of configurations $(C,C')$.  The general form of these trajectory observables is,
\begin{equation}
A(\omega_t) = \sum_\mu \alpha_\mu n_\mu(\omega_t) .
\label{A}
\end{equation}
In the MPS formalism, such observables can be expressed in terms of operators acting on the output state,
\begin{equation}
\A = \sum_\mu \int_0^t dt' \, \alpha_\mu \, a_\mu^\dagger(t') a_\mu(t') ,
\label{AA}
\end{equation}
since $\A | \omega_t \rangle = A(\omega_t) | \omega_t \rangle$. 

One is typically interested in the probability distribution $P_t(A)$ of $A$ over dynamics up to time $t$.  In terms of the cMPS state that encodes the ensemble of trajectories this probability is,
\begin{equation}
P_t(A) = \langle -_{\rm tot} | \delta( A - \A ) | \Psi_t \rangle .
\label{PA}
\end{equation}
The associated moment generating function (MGF) then reads,
\begin{equation}
Z_t(s) = \sum_A e^{-s A} P_t(A) 
=
\langle -_{\rm tot} | e^{-s \A} | \Psi_t \rangle .
\label{Z}
\end{equation}
We can now define a {\em tilted} cMPS (cf.\ \cite{Lesanovsky2013,Kiukas2015}) by application of the $e^{-s \A}$ operator, $| \Psi_t(s) \rangle = e^{-s \A} | \Psi_t \rangle$.  From Eqs.\ (\ref{MPS1},\ref{MPS2}) we see that this new cMPS is given by 
\begin{align}
| \Psi_t(s) \rangle =
\T 
\exp \left\{ 
\int_0^t dt' 
\left[ 
\sum_{\mu=1}^{\NJ} e^{-s \alpha_\mu} \W_\mu \otimes a_\mu^\dagger(t') 
\nonumber
\right. \right. \\
\left. \left.
\phantom{\sum_{\mu=1}^{\NJ}}
- \R \otimes \Iout 
\right]
\right\}
| C_0 \rangle \otimes | \Omega \rangle .
\label{MPS}
\end{align}
The tilted state $| \Psi_t(s) \rangle$ encodes a biased, or {\em conditioned}, ensemble of trajectories, i.e., one where trajectories are conditioned to have an average value of $A$ that is set by the parameter $s$ \cite{Lecomte2007,Garrahan2007,Chetrite2015}.  This conditioned ensemble can be termed ``canonical'' \cite{Chetrite2013}, in analogy with the equilibrium canonical ensemble where the conditioning on configurations is on the average energy.  

The norm of $| \Psi_t(s) \rangle$ is the MGF, as seen from \er{Z},
\begin{equation}
\langle -_{\rm tot} | \Psi_t(s) \rangle = Z_t(s) .
\label{Ns}
\end{equation}
For long times we expect the MGF $Z_t(s)$ and probability $P_t(A)$ to have large-deviation forms \cite{Ruelle2004,Gaspard2005,Touchette2009}, 
\begin{equation}
Z_t(s) \approx e^{t \theta(s)} 
\; , \;\;\;
P_t(A) \approx e^{-t \varphi(A/t)} ,
\label{LD}
\end{equation}
with the scaled cumulant generating function $\theta(s)$ and the rate function $\varphi(a)$ related by a Legendre transform, 
\begin{equation}
\varphi(a) = -\min_s \left[ \theta(s) + s \, a \right] ,
\label{LT}
\end{equation}
that is, 
\begin{equation}
\varphi(a) = - \theta[s(a)] - s(a) \, a ,
\end{equation}
with $s$ and $a$ related through 
\begin{equation}
a(s) = -\theta'(s) .
\label{as}
\end{equation}
This structure is analogous to that of equilibrium statistical mechanics: $A$ is an order parameter and $s$ its conjugate field; the large size limit is given by large $t$; $P_t(A)$ is the order parameter distribution, with $\varphi(A/t)$ playing the role of an entropy density; $Z_t(s)$ is a ``partition sum'', with $\theta(s)$ playing the role of a free-energy density for trajectories.

By tracing out the output in the tilted state $| \Psi_t(s) \rangle$ we see that the conditioned ensemble is generated by a {\em tilted generator} \cite{Touchette2009},
\begin{equation}
\langle -_{\rm out} | \Psi_t(s) \rangle = e^{t \L(s)} | C_0 \rangle ,
\label{NS2}
\end{equation}
where 
\begin{equation}
\L(s) = \sum_{\mu=1}^{N_{\rm J}} e^{-s \alpha_\mu} \W_{\mu} - \R .
\label{Ls}
\end{equation}
The operator $\L(s)$ is useful not only because it generates the conditioned ensemble, but also because its largest eigenvalue is $\theta(s)$.  

\smallskip

Note the following:

\smallskip $\bullet$ 
The ensemble of trajectories encoded by $| \Psi_t(s) \rangle$, sometimes termed $s$-ensemble \cite{Hedges2009}, is canonical in the sense that trajectories are conditioned on the average of an observable \cite{Chetrite2013}.  We can also consider a ``microcanonical'' ensemble of trajectories conditioned on a fixed value of $A$ \cite{Chetrite2013}.  The corresponding system-output state that encodes it is
\begin{equation}
| \Psi_t^A \rangle = \delta( A - \A ) | \Psi_t \rangle .
\label{MPSA}
\end{equation}
The norm of this state is $P_t(A)$, see \er{PA}.  In contrast with canonical conditioning, in general there is no generator associated with this ensemble [as the $\delta( A - \A )$ operator cannot be easily ``sliced'' into the time integral of \er{MPS2}].

\smallskip $\bullet$ 
Up to now we have considered a single observable $A$, but the definitions above are easily generalised to multiple observables, $\vec{A}=(A_1,A_2,\ldots)$, and multiple conjugate fields, $\vec{s}=(s_1,s_2,\ldots)$, with $| \Psi_t(\vec{s}) \rangle = e^{- \vec{s}\cdot \vec{\A}} | \Psi_t \rangle$, where $\vec{\A}=(\A_1,\A_2,\ldots)$ are the corresponding output operators.  

\smallskip $\bullet$ 
We can also consider observables that depend on the configuration of the system between jumps, 
\begin{equation}
B(\omega_t) = \int_0^t dt' \beta[C(t')] .
\label{B}
\end{equation}
The associated cMPS for the ensemble of trajectories conditioned on the average of such an operator, $| \Psi_t(s) \rangle$, has the form of \er{MPS2}, but with $\R \to \R + s \, \B$, where 
\begin{equation}
\B = \sum_C \beta(C) |C \rangle \langle C| .
\label{BB}
\end{equation}

\subsection{Example}
\label{Exx}

For an illustration of the above, lets return to the example of Subsection \ref{Ex}.  The simplest conditioning would be on the total number of hops $K$ in a trajectory.  From \er{ExPo} it follows---since the sum over all trajectories is just the integration over all jump times---that $K$ is Poisson distributed, as expected, 
\begin{equation}
P_{t}(K) = e^{-\gamma t} \frac{(\gamma t)^{K}}{K!} . 
\label{PKay}
\end{equation}
The corresponding tilted cMPS, \er{MPS}, reads
\begin{equation}
| \Psi_t(s) \rangle =
e^{-\gamma t} \;
\T 
e^{\int_0^t dt' e^{-s} \W \otimes a^\dagger(t')}
| x_0 \rangle \otimes | \Omega \rangle .
\label{MPSex}
\end{equation}
The norm of this cMPS, $Z_{t}(s)$, is obtained immediately by observing that the factor $e^{-s}$ can be absorbed into $\W$ by a redefinition of the rate $\gamma$, giving,
\begin{equation}
Z_{t}(s) = e^{t \gamma \left(e^{-s}-1\right)} .
\end{equation}
Note that the MGF of $K$ is exponential in time for all times, $Z_{t}(s) = e^{t \theta(s)}$, where
\begin{equation}
\theta(s) = \gamma \left(e^{-s}-1\right) ,
\label{tss}
\end{equation}
is also the largest eigenvalue of the tilted generator, \er{Ls},
\begin{equation}
\L(s) = e^{-s} \gamma \sum_{x} |x+1\rangle \langle x| - \gamma \sum_{x} |x \rangle \langle x| .
\label{Lss}
\end{equation}
The LD form of $Z_{t}(s)$ is therefore valid for all times, not just asymptotically.  This is a consequence of $K$ being Poissonian: a Poisson process conditioned on its average activity is simply another Poisson process with modified rates (see discussion below).

A less trivial conditioning is one based on a non-uniform weighting of jumps.  An example would be to consider as a trajectory observable the total number $K_{\rm e}$ of jumps to even sites.  For such conditioning the corresponding tilted operator reads 
\begin{align}
\L_{\rm e}(s) = e^{-s} \gamma \sum_{x={\rm odd}} |x+1\rangle \langle x|
+ \gamma \sum_{x={\rm even}} |x+1\rangle \langle x|
\nonumber \\
 - \gamma \sum_{x} |x \rangle \langle x| .
\label{Lsoe}
\end{align}
This operator is easy to diagonalise (in the following we assume $L$ is even for simplicity---see below for the full spectral analysis).  Its largest eigenvalue corresponds to the scaled cumulant generating function, \er{LD}, 
\begin{equation}
\theta_{\rm e}(s) = \gamma \left( e^{-s/2} - 1 \right) .
\label{tsoe}
\end{equation}
From \er{tsoe} it follows that the statistics of $K_{\rm e}$ is sub-Poissonian: while its average is $\langle K_{\rm e} \rangle = t \gamma/2$ as expected (as we are counting only half of the jumps), its variance is $\langle K_{\rm e}^2 \rangle- \langle K_{\rm e} \rangle^2 = t \gamma/4 < \langle K_{\rm e} \rangle$.  As we will see below, the conditioned ensemble associated to \er{Lsoe} is not related to a simple rescaling of rates. 

The cMPSs conditioned on the exact number of $K$ or $K_{\rm e}$ are easy to construct.  When conditioning on the total number of jumps $K$, the cMPS is a single layer of the bosonic Fock space,
\begin{align}
| \Psi_t^K \rangle 
=
e^{- t \gamma}
\int_0^t dt_1 
\cdots
\int_{t_{K-1}}^t dt_K \, \W^{K} | x_0 \rangle
\nonumber \\
\otimes \, a^{\dagger}(t_K) \cdots a^{\dagger}(t_{1}) \,
| \Omega \rangle .
\end{align}
The norm of this state is \er{PKay}. In contrast, when conditioning on odd-to-even jumps $K_{\rm e}$ one gets the superposition of two layers,
\begin{align}
| \Psi_t^{K_{\rm e}} \rangle 
=
e^{- t \gamma}
\int_0^t dt_1 
\cdots
\int_{t_{2K_{\rm e}-1}}^t dt_{2K_{\rm e}} \, \W^{2K_{\rm e}} | x_0 \rangle
\nonumber \\
\otimes \, a^{\dagger}(t_{2K_{\rm e}}) \cdots a^{\dagger}(t_{1}) \,
| \Omega \rangle
\nonumber \\
+
e^{- t \gamma}
\int_0^t dt_1 
\cdots
\int_{t_{2K_{\rm e}}}^t dt_{2K_{\rm e} \pm 1} \, \W^{2K_{\rm e} \pm 1} | x_0 \rangle
\nonumber \\
 \otimes \, a^{\dagger}(t_{2K_{\rm e} \pm 1}) \cdots a^{\dagger}(t_{1}) \,
| \Omega \rangle ,
\label{mpskp}
\end{align}
where $\pm$ depends on whether the starting position $x_{0}$ is even or odd, respectively.  The norm of this state gives the probability of $K_{\rm e}$, 
\begin{equation}
P_{t}(K_{\rm e}) = e^{-\gamma t} \frac{(\gamma t)^{2K_{\rm e}}}{(2K_{\rm e})!}
+ e^{-\gamma t} \frac{(\gamma t)^{2K_{\rm e} \pm 1}}{(2K_{\rm e} \pm 1)!} , 
\label{PKayp}
\end{equation}
which is sub-Poissonian, in agreement with \er{tsoe}.

\section{Gauge invariance and driven process as gauge fixing}

A cMPS state such as \er{MPS2} or \er{MPS} is invariant under the following gauge transformations,
\begin{equation}
\tW_\mu = \G \, \W_\mu \, \G^{-1} 
\; , \;\;\;
\tR = \G \, \R \, \G^{-1} - \frac{\partial}{\partial t'} \G \, \G^{-1},
\label{G}
\end{equation}
where $\G$ are invertible, and possibly time-dependent, $D \times D$ real matrices.   This gauge invariance is directly analogous to that of quantum cMPSs \cite{Haegeman2013}.  The so-called {\em driven} process \cite{Chetrite2015} associated to a canonically conditioned ensemble follows immediately from this gauge invariance of cMPSs as we see now.

The canonically conditioned ensemble is generated by the operator $\L(s)$, but this tilted operator is not stochastic: in general its largest eigenvalue is non-vanishing and the associated left eigenvector is not the flat state $\fs$.  A natural question is whether there is an alternative non-conditioned process which generates the same conditioned ensemble of trajectories but which is properly stochastic: this is the driven process (alternatively called ``auxiliary process'' \cite{Jack2010}).  
As discussed in \cite{Chetrite2015}, the generator of the driven process can be obtained from the tilted operator $\L(s)$ through a (generalised) Doob transform.  We now show that such a transform corresponds to a particular choice of gauge in the cMPS formalism.  

We wish to find a gauge transformation \er{G} that maps the (normalised) conditioned cMPS state encoding the conditioned ensemble to the cMPS of some other non-conditioned and properly stochastic process.  The normalised conditioned cMPS is
\begin{equation}
| \Psi_t^{{\rm norm}}(s) \rangle = \frac{1}{Z_t(s)} \, | \Psi_t(s) \rangle .
\label{snorm}
\end{equation}
We can absorb the normalisation by defining $Z_t(s) = e^{\int_{0}^{t} f(t') dt'}$, so that $f(t') = \frac{\partial}{\partial t'} \log Z_{t'}(s)$.  The normalised conditional cMPS then reads
\begin{align}
| \Psi_t^{{\rm norm}}(s) \rangle =
\T 
\exp \left\{ 
\int_0^t dt' 
\left[ 
\sum_{\mu=1}^{\NJ} e^{-s \alpha_\mu} \W_\mu \otimes a_\mu^\dagger(t') 
~~~~~
\label{MPSn}
\right. \right. \\
\left. \left.
\phantom{\sum_{\mu=1}^{\NJ}}
- [\R + f(t') ] \otimes \Iout 
\right]
\right\}
| C_0 \rangle \otimes | \Omega \rangle .
\nonumber
\end{align}
Note that this is the same as \er{MPS} but with the replacement,
\begin{equation}
\R \to \R + f(t') . 
\label{Rf}
\end{equation}

We aim to find
\begin{equation}
| \Psi_t^{{\rm norm}}(s) \rangle = | \tilde{\Psi}_t \rangle ,
\label{c2d}
\end{equation}
where 
\begin{equation}
| \tilde{\Psi}_t \rangle =
\T 
e^{ 
\int_0^t dt' 
\left[ 
\sum_{\mu=1}^{\NJ} \tW_\mu(t') \otimes a_\mu^\dagger(t') 
- \tR(t') \otimes \Iout 
\right]
}
| C_0 \rangle \otimes | \Omega \rangle ,
%\label{MPSd}
\nonumber
\end{equation}
such that $\tL_{t'} = \sum_\mu \tW_\mu(t') - \tR(t')$, which in general can be time dependent, is a stochastic operator, 
\begin{equation}
\langle - | \tL_{t'} = 0 .
\label{ts}
\end{equation}

Combining \er{ts} with \er{G} then gives the equation that the gauge transformation $\G$ must obey to transform the conditioned process into the driven one, \er{c2d},
\begin{equation}
\frac{\partial}{\partial t'} \langle - | \G_{t'} = \langle - | \G_{t'} \, [f(t') - \L(s)] .
\label{eqG}
\end{equation} 
If we assume that $\G$ is a diagonal operator, and that 
the boundary condition for \er{eqG} at the final time $t$ is $\langle - | \G_t = \langle - |$, then \er{eqG} formally integrates to give, 
\begin{equation}
\G_{t'} = \sum_{C} \frac{\langle - | e^{(t-t') \L(s)} | C \rangle}{Z_{t-t'}(s)} \, | C \rangle \langle C | .
\label{Gsol}
\end{equation} 
Note that since $\G_{0}$ is non-trivial, strictly speaking the transformation \er{G} includes also a transformation of the initial condition $| C_0 \rangle \to \G_{0} | C_0 \rangle$.

When the overall trajectory length $t$ is large compared to intermediate times $t'$ the evolution operator $e^{(t-t') \L(s)}$ is exponentially dominated by its largest eigenvalue.  That is, $e^{(t-t') \L(s)} \approx e^{(t-t') \theta(s)} | r_{s} \rangle \langle l_{s} |$, where $| r_{s} \rangle$ and $\langle l_{s} |$ are the right and left eigenvectors of $\L(s)$ 
\begin{equation} 
\L(s) \, | r_{s} \rangle = \theta(s) \, | r_{s} \rangle 
\; , \;\;\;
\langle l_{s} | \, \L(s) = \theta(s) \, \langle l_{s} | , 
\label{rl}
\end{equation}
Where the normalisation is $\langle l_{s} |r_{s} \rangle = \langle - |r_{s} \rangle = 1$. Similarly, the denominator in \er{Gsol} simplifies to $Z_{t-t'}(s) \approx e^{(t-t') \theta(s)}$, cf.\ \er{LD}. 
This means that for trajectories where the overall time $t$ is large we have $\G_{t'} \approx \G_{\infty}$, where
\begin{equation}
\G_{\infty} = \langle l_{s} | C_0 \rangle^{-1} 
\sum_{C} \langle l_{s} | C \rangle \, | C \rangle \langle C | 
.
\label{Ginf}
\end{equation}
This is in fact the asymptotic solution to the generalised Doob transform from the conditioned to driven process given in Refs.\ \cite{Jack2010,Garrahan2010,Chetrite2015}, that of a diagonal matrix whose entries are the coefficients of the leading left eigenvector of $\L(s)$ [the prefactor that depends on the initial conditions is irrelevant as it drops out of \er{G}].

The gauge transformation \er{Gsol} for arbitrary $t$ and $t' \in [0,t]$ obtained here in the ``Schrodinger picture'' (where what is propagated in time is the probability) is the generalised Doob transform that gives the driven process first found in \cite{Chetrite2015} in the ``Heisenberg'' picture (where what are propagated are observables).  Note that the relation between generalised Doob transforms and gauge transformations was also considered in \cite{Chetrite2011}.

\subsection{Example}
\label{Exxx}

We now illustrate how to obtain the driven process with the example of Subsections \ref{Ex} and \ref{Exx}.  

For the case of conditioning on the average value of the total number of jumps $K$, where the tilted generator is \er{Lss}, the mapping between the conditioned and driven process is trivial.  The driven process is obtained simply by normalising the cMPS, which amounts to shifting the escape rate $\R$ by the scaled cumulant generating function $\theta(s)= \gamma \left(e^{-s}-1\right)$, cf.~\er{MPSn}, since the function $f(t')=\theta(s)$ is time independent.  The gauge transformation in this case is trivial and $\G_{t'}$ is the identity for all times.  The transformed generator is time independent a and reads, 
\begin{equation}
\tilde{\L}(s) = e^{-s} \gamma \sum_{x} |x+1\rangle \langle x| - e^{-s} \gamma \sum_{x} |x \rangle \langle x| ,
\label{tLss}
\end{equation}
which corresponds to the stochastic generator of the original problem with a change of rates $\gamma \to e^{-s} \gamma$.  As anticipated, this shows that the ensemble of trajectories conditioned on the average of the total number of jumps is the same as the ensemble of trajectories of the same problem with scaled rates.  

The case where conditioning is on the average of the number of odd-to-even jumps $K_{\rm e}$ is less trivial.  In order to explicitly solve the gauge transformation \er{Gsol} we need to diagonalise $\L_{\rm e}(s)$, which is readily done exploiting translation invariance.  The eigenvalues of $\L_{\rm e}(s)$ are
\begin{equation}
\lambda_{q}^{\pm} = \gamma \left(\pm e^{-\frac{s}{2}+i q} - 1 \right) ,
\label{es}
\end{equation}
with
\begin{equation}
q=0,\frac{2\pi}{L},\frac{4\pi}{L},\cdots,\frac{2\pi(L/2-1)}{L} .
\end{equation}
From the above we see that $\lambda_{0}^{+} = \theta_{\rm e}(s)$ of \er{tsoe}.  The corresponding right eigenvectors are
\begin{equation*}
| r_{q}^{\pm} \rangle = \frac{1}{L}
\left(
\begin{array}{ccccc}
\pm e^{s/2} & 0 & \cdots & 0 & 0 \\
0 & 1 & \cdots & 0 & 0 \\
\vdots & \vdots & \ddots & \vdots & \vdots \\
0 & 0 & \cdots & \pm e^{s/2} & 0 \\
0 & 0 & \cdots & 0 & 1 \\
\end{array}
\right)
\cdot
\left(
\begin{array}{c}
e^{i q (L-1)} \\
e^{i q (L-2)} \\
\vdots \\
e^{i q} \\
1
\end{array}
\right) ,
\end{equation*}
while the left eigenvectors read
\begin{equation*}
| l_{q}^{\pm} \rangle =
\left(
\begin{array}{ccccc}
\pm e^{- s/2} & 0 & \cdots & 0 & 0 \\
0 & 1 & \cdots & 0 & 0 \\
\vdots & \vdots & \ddots & \vdots & \vdots \\
0 & 0 & \cdots & \pm e^{- s/2} & 0 \\
0 & 0 & \cdots & 0 & 1 \\
\end{array}
\right)
\cdot
\left(
\begin{array}{c}
e^{i q} \\
e^{i 2 q} \\
\vdots \\
e^{i (L-1) q} \\
1
\end{array}
\right) ,
\end{equation*}
with normalisation $\langle l_{q}^{\sigma} | r_{q'}^{\sigma'} \rangle = \delta_{\sigma \sigma'} \delta_{q q'}$.  In this notation the eigenvectors of the largest eigenvalue are
\begin{equation*}
|r_{s}\rangle = |r_{0}^{+}\rangle
= \frac{1}{L}
\left(
\begin{array}{c}
e^{s/2} \\
1 \\
\vdots \\
e^{s/2} \\
1
\end{array}
\right) , \;\;
|l_{s}\rangle = |l_{0}^{+}\rangle
= 
\left(
\begin{array}{c}
e^{-s/2} \\
1 \\
\vdots \\
e^{-s/2} \\
1
\end{array}
\right) .
\end{equation*}

From the spectral decomposition or $\L_{\rm e}(s)$ we can write the evolution operator as,
\begin{equation}
e^{t \L_{\rm e}(s)} = \sum_{q} 
\left(
e^{t \lambda_{q}^{+}} |r_{q}^{+} \rangle \langle l_{q}^{+} |
+
e^{t \lambda_{q}^{-}}
l_{q}^{-} |r_{q}^{-} \rangle \langle l_{q}^{-} |
\label{eL}
\right) ,
\end{equation}
and calculate the MGF,
\begin{equation}
Z_{{\rm e},t}^{\pm}(s) = e^{-\gamma t} \cosh{\left(\gamma t e^{-\frac{s}{2}}\right)} + 
e^{-\gamma t \pm \frac{s}{2}} \sinh{\left(\gamma t e^{-\frac{s}{2}}\right)} .
\label{ZZtt}
\end{equation}
Again the $\pm$ depends on whether the starting position is even or odd, respectively.  At long times the MGF is indeed of the form of \er{LD} with rate \er{tsoe}.  With \ers{eL}{ZZtt} we can find the explicit form of the gauge transformation operator \er{Gsol},
\begin{equation}
\G_{t'} = \frac{Z_{{\rm e},t-t'}^{+}(s)}{Z_{{\rm e},t-t'}^{-}(s)} \sum_{x = {\rm even}} |x \rangle \langle x |
+  \sum_{x = {\rm odd}} |x \rangle \langle x | ,
\label{Gtp}
\end{equation} 
where we have assumed that the initial position $x_{0}$ is an odd site, say $x_{0}=1$.  From this we find that the driven process has a stochastic generator which is time dependent, cf.~\ers{G}{c2d}, given by,
\begin{align}
\tilde{\L_{\rm e}}(s) &= \G \, \L_{\rm e} \, \G^{-1} - f_{\rm e}(t') 
+\frac{\partial}{\partial t'} \G \, \G^{-1}
\nonumber \\
&= \sum_x \tilde{\W}_{{\rm e},x}(t') - \tilde{\R}_{{\rm e}}(t') ,
\label{Ltp}
\end{align}
where 
\begin{align}
\tilde{\W}_{{\rm e},x}(t') 
&=
\left\{
\begin{array}{rcl}
e^{-s} \frac{Z_{{\rm e},t-t'}^{+}(s)}{Z_{{\rm e},t-t'}^{-}(s)} \, |x + 1 \rangle \langle x |
& & x = {\rm even} \\
\\
\frac{Z_{{\rm e},t-t'}^{-}(s)}{Z_{{\rm e},t-t'}^{+}(s)} \, |x + 1 \rangle \langle x |
& & x = {\rm odd} \\
\end{array}
\right. 
\label{tWp}
\\
\tilde{\R}_{{\rm e}}(t') &= \sum_x \tilde{\W}_{{\rm e},x}(t') |x \rangle \langle x | ,
\label{tRp}
\end{align}
and $f_{{\rm e}}(t') = \partial_{t'} \log Z_{{\rm e},t'}(s)$.
This means that the driven process generated by $\tilde{\L_{\rm e}}(s)$ is a proper stochastic one of unidirectional hopping dynamics, as in the original process \er{Ex3}, but where the rates alternate between the two possibilities of \er{tWp}.  

\smallskip

Note the following:

\smallskip $\bullet$ 
The driven generator \ers{Ltp}{tRp} is time dependent for arbitrary time $t'$, and in general is not a simple rescaling of the original process, \er{Ex3}, as odd and even rates are different.  This contrasts with the driven generator when the conditioning is on the total number of jumps, \er{tLss}, which is simply the original process with $\gamma \to e^{-s} \gamma$.

\smallskip $\bullet$ 
In the limit of large $t$ with $t' \ll t$, from \er{Gtp} we get for the transformation operator,
\begin{equation}
\G_{t'} \to \G_{\infty} \propto \sum_{x = {\rm even}} |x \rangle \langle x |
+  \sum_{x = {\rm odd}} e^{-s/2} |x \rangle \langle x | ,
\label{Gtp}
\end{equation} 
in accordance with \er{Ginf}.  Only in this limit the driven process coincides with a rescaling of the original dynamics, as $\tilde{\L_{\rm e}}(s)$ of \er{Ltp} becomes of the form \er{tLss}.

\bigskip

\section{Equivalence of ensembles}

Using the cMPS formalism it is also easy to prove the equivalence of trajectory ensembles \cite{Chetrite2013}.  Consider for example the conditioned trajectory ensemble encoded in the normalised cMPS $| \Psi_t^{{\rm norm}}(s) \rangle $, cf.~\er{snorm}, where conditioning is through the average of the trajectory observable $\langle A \rangle(s)$, as compared to that encoded in the normalised cMPS $| \Psi_t^{A,{\rm norm}} \rangle = P_{t}^{-1}(A) \, | \Psi_t^{A} \rangle$ where conditioning is on the actual value of $A$.  One expects that these ``canonical'' and ``microcanonical'' trajectory ensembles should be equivalent at long times for some appropriate value of $s$ such that $\langle A \rangle_{\rm cano}(s) = A_{\rm micro}$ \cite{Chetrite2013}.  Since the two ensembles are fully encoded in two real vectors in the system-output space we can prove this equivalence by considering the distance between these vectors.  

For simplicity we focus first on the case where the trajectory observable is the total number of transitions in a trajectory, that is, the dynamical activity $K$.  Going back to the definitions \ers{Pomega}{MPS1} we can write the two cMPS vectors as,
\begin{widetext}
\begin{align}
| \Psi_t^{{\rm norm}}(s) \rangle 
&=
\frac{1}{Z_{t}(s)} \,
\sum_{K=0}^{\infty} e^{{-sK}} \sum_{\mu_{1} \cdots \mu_{K}=1}^{\NJ}
\int_{0 \leq t_1 \cdots t_K \leq t}
\V \left( \omega_{t}^{\mu_{1}t_{1}\cdots\mu_{K}t_{K}} \right) 
| C_{0} \rangle \otimes | \omega_{t}^{\mu_{1}t_{1}\cdots\mu_{K}t_{K}} \rangle ,
\label{cn}
\\
| \Psi_t^{K,{\rm norm}} \rangle 
&=
\frac{1}{P_{t}(K)} \,
\sum_{\mu_{1} \cdots \mu_{K}=1}^{\NJ}
\int_{0 \leq t_1 \cdots t_K \leq t}
\V \left( \omega_{t}^{\mu_{1}t_{1}\cdots\mu_{K}t_{K}} \right) 
| C_{0} \rangle \otimes | \omega_{t}^{\mu_{1}t_{1}\cdots\mu_{K}t_{K}} \rangle ,
\label{mcn}
\end{align}
\end{widetext}
where by $\omega_{t}^{\mu_{1}t_{1}\cdots\mu_{K}t_{K}}$ we are denoting the trajectory with jumps $[(\mu_1,t_{1}), (\mu_2,t_{2}), \cdots, (\mu_{K},t_{K})]$, and $| \omega_{t}^{\mu_{1}t_{1}\cdots\mu_{K}t_{K}} \rangle$ stands for 
\begin{equation}
| \omega_{t}^{\mu_{1}t_{1}\cdots\mu_{K}t_{K}} \rangle = 
a_{\mu_{K}}^{\dagger}(t_{K}) \cdots a_{\mu_{2}}^{\dagger}(t_{2}) \,
a_{\mu_{1}}^{\dagger}(t_{1}) \,
| \Omega \rangle .
\label{oma}
\end{equation}

We will prove two forms of equivalence.  The first and simplest one is the {\em equivalence of concentration} \cite{Chetrite2013,Kiukas2015}: for long times, the canonical and microcanonical trajectory measures concentrate exponentially on the same region of their support, as long as $K$ and $s$ are related by \er{as}.  
This can be expressed in terms of the the ratio of the probabilities of a trajectory in the ensemble conditioned by $s$ and in the ensemble conditioned by $K$.  For long times, when $K$ and $s$ are connected through \er{as}, this ratio is sub-exponential in time \cite{Chetrite2013}.
In the MPS formalism this form of equivalence can be proven by considering 
the distance between the vectors that encode the two ensembles, $\Vert \, | \Psi_t^{{\rm norm}}(s) \rangle - | \Psi_t^{K,{\rm norm}} \rangle \, \Vert_{1}$.  
Moreover, since each of the vectors has unit norm, 
$\Vert \, | \Psi_t^{{\rm norm}}(s) \rangle \Vert_{1} = \Vert \, | \Psi_t^{K,{\rm norm}} \rangle \, \Vert_{1} = 1$, the distance between them is bounded by $2$, so that the quantity $2 - \Vert \, | \Psi_t^{{\rm norm}}(s) \rangle - | \Psi_t^{K,{\rm norm}} \rangle \, \Vert_{1}$ plays the role of an ``overlap'' between the probabilities (cf.~the quantum case \cite{Kiukas2015} where the overlap is actually the internal product between the cMPS vectors). The output states $| \omega_{t} \rangle$ are orthonormal, and both $| \Psi_t^{{\rm norm}}(s) \rangle$ and $| \Psi_t^{K,{\rm norm}} \rangle$ are real positive vectors, so we get
\begin{widetext}
\begin{align}
2-
\Vert \, | \Psi_t^{{\rm norm}}(s) \rangle - | \Psi_t^{K,{\rm norm}} \rangle \, \Vert_{1} 
=&
\,
2-
\sum_{K' \neq K} \frac{e^{{-sK'}}}{Z_{t}(s)} 
\sum_{\mu_{1} \cdots \mu_{K'}}
\int_{0 \leq t_1 \cdots t_{K'} \leq t}
\langle -_{\rm tot} | 
\V \left( \omega_{t}^{\mu_{1}t_{1}\cdots\mu_{K'}t_{K'}} \right) 
| C_{0} \rangle 
\nonumber
\\
&
- 
\left| \frac{e^{{-sK}}}{Z_{t}(s)} - \frac{1}{P_{t}(K)} \right|
\sum_{\mu_{1} \cdots \mu_{K}}
\int_{0 \leq t_1 \cdots t_K \leq t}
\langle -_{\rm tot} | 
\V \left( \omega_{t}^{\mu_{1}t_{1}\cdots\mu_{K}t_{K}} \right) 
| C_{0} \rangle 
\nonumber \\
=&
\, 
1 + 
\frac{e^{{-sK}}}{Z_{t}(s)} 
\sum_{\mu_{1} \cdots \mu_{K}}
\int_{0 \leq t_1 \cdots t_{K} \leq t}
\langle -_{\rm tot} | 
\V \left( \omega_{t}^{\mu_{1}t_{1}\cdots\mu_{K}t_{K}} \right) 
| C_{0} \rangle 
\nonumber
\\
&
- 
\left| \frac{e^{{-sK}}}{Z_{t}(s)} - \frac{1}{P_{t}(K)} \right|
\sum_{\mu_{1} \cdots \mu_{K}}
\int_{0 \leq t_1 \cdots t_K \leq t}
\langle -_{\rm tot} | 
\V \left( \omega_{t}^{\mu_{1}t_{1}\cdots\mu_{K}t_{K}} \right) 
| C_{0} \rangle 
\nonumber
\\
=& \,
\frac{2 e^{{-sK}} P_{t}(K) }{Z_{t}(s)} ,
\label{cmc}
\end{align}
\end{widetext}
where the final simplification is because the summations in the third and fourth lines give $P_{t}(K)$ and we have that $ e^{{-sK}} P_{t}(K) \leq Z_{t}(s) = \sum_{K'} e^{-sK'} P_{t}(K')$.  At long times we can write \er{cmc} using the large-deviation forms \er{LD}, so that,
\[
2- \Vert \, | \Psi_t^{{\rm norm}}(s) \rangle - | \Psi_t^{K,{\rm norm}} \rangle \, \Vert_{1} \approx
2 e^{{t \left[ \theta(s) - s \frac{K}{t} - \varphi\left(\frac{K}{t}\right) \right] }} .
\]
If we minimise with respect to $s$ for fixed $K$ we obtain from \er{LT} that
\begin{equation}
2- \Vert \, | \Psi_t^{{\rm norm}}(s) \rangle - | \Psi_t^{K,{\rm norm}} \rangle \, \Vert_{1} = e^{o(t)} ,
\label{equiv}
\end{equation}
as long as $s$ is such that, cf.~\ers{LT}{as},
\begin{equation}
\frac{K}{t} = - \theta'(s) .
\label{sstar}
\end{equation}
The equality \er{equiv} is correct up to corrections that are sub-exponential in time.  This way of proving ensemble equivalence in terms of the distance between the corresponding cMPS vectors is the classical analog, for real vectors in an L1-space, of that of Ref.\ \cite{Kiukas2015} for the case of quantum stochastic dynamics where the cMPS vectors are complex and belong to an L2-space (the system-output Hilbert space). 

A more detailed form of equivalence, termed {\em operational equivalence} in Ref.~\cite{Kiukas2015}, can be proved in the following way.  The idea is to 
consider a conditioned ensemble where the conditioning is on the whole time extension $t$, with $t$ large, but focus only on the portion of trajectories up to time $t_0 \ll t$.  We now ask the question: how does the distribution of sub-trajectories up to time $t_0$, from an ensemble conditioned microcanonically by some time-extensive quantity $A$ for all $t$, compare to that for the same sub-trajectories of an ensemble conditioned canonically by $s$.  This can be answered by comparing the ``reduced'' cMPSs where all information after $t_0$ is traced over.  

In order to construct the reduced states we first need to partition trajectories at time $t_0$:  
\begin{equation}
\omega_{t}^{\mu_{1}t_{1}\cdots\mu_{K}t_{K}} = \omega_{t_0}^{\mu_{1}t_{1}\cdots\mu_{K_0}t_{K_0}} \times
\omega_{t-t_0}^{\mu_{K_0+1}t_{K_0+1}\cdots\mu_{K}t_{K}} .
\label{oo}
\end{equation}
Here $\omega_{t_0}^{\mu_{1}t_{1}\cdots\mu_{K_0}t_{K_0}} \equiv \omega_{t_0}$ indicates the sequence of $K_0$ jumps that occur between time $0$ and $t_0$, and $\omega_{t-t_0}^{\mu_{K_0+1}t_{K_0+1}\cdots\mu_{K}t_{K}} \equiv \omega_{+}$ the sequence of the $K-K_0$ jumps occurring after $t_0$. The corresponding output states are similarly split,
\begin{equation}
| \omega_{t} \rangle = | \omega_{t_0} \rangle \otimes | \omega_{+} \rangle ,
\label{ooo}
\end{equation}
and the action on the system can be written as the product, cf.~\er{VV},
\begin{equation}
\V \left( \omega_{t} \right) = 
\V \left( \omega_{+} \right) 
\V \left( \omega_{t_0} \right) .
\end{equation}
We also define the trace state over all the output after time $t_0$,
$\langle -_{{\rm out}_+} | = \sum_{\omega_+} \langle \omega_+ |$, cf.~\er{N2}.
The normalised cMPSs, cf.~\era{cn}{mcn}, for conditioning on arbitrary trajectory observable, cf.~\era{A}{B}, are
\begin{widetext}
\begin{align}
| \Psi_t^{{\rm norm}}(s) \rangle 
&=
\frac{1}{Z_{t}(s)} \,
\sum_{K=0}^{\infty} \sum_{\mu_{1} \cdots \mu_{K}=1}^{\NJ}
\int_{0 \leq t_1 \cdots t_K \leq t}
e^{-s \A} \,
\V \left( \omega_{t}^{\mu_{1}t_{1}\cdots\mu_{K}t_{K}} \right) 
| C_{0} \rangle \otimes | \omega_{t}^{\mu_{1}t_{1}\cdots\mu_{K}t_{K}} \rangle ,
\label{acn}
\\
| \Psi_t^{A,{\rm norm}} \rangle 
&=
\frac{1}{P_{t}(A)} \,
\sum_{K=0}^{\infty} \sum_{\mu_{1} \cdots \mu_{K}=1}^{\NJ}
\int_{0 \leq t_1 \cdots t_K \leq t}
\delta(A - \A) \,
\V \left( \omega_{t}^{\mu_{1}t_{1}\cdots\mu_{K}t_{K}} \right) 
| C_{0} \rangle \otimes | \omega_{t}^{\mu_{1}t_{1}\cdots\mu_{K}t_{K}} \rangle .
\label{amcn}
\end{align}
Here $\A$ is the operator acting on the output state that gives the value of the trajectory observable, $\A | \omega_t \rangle = A(\omega_t) | \omega_t \rangle$.  Since $A$ is time extensive, the operator $\A$ can be split as a sum of two commuting operators, $\A = \A_0 + \A_+$, where $\A_0$ acts on the first part of the output, $\A_0 | \omega_{t_0} \rangle = A(\omega_{t_0}) | \omega_{t_0} \rangle$, and $\A_+$ on the second part, $\A_+ | \omega_{+} \rangle = A(\omega_{+}) | \omega_{+} \rangle$. Tracing the canonical cMPS, \er{acn}, over the output after $t_0$ and over the system gives the reduced state,
\begin{align}
| \psi_{t_0}(s) \rangle 
&= 
\langle - | \otimes \langle -_{{\rm out}_+} | \Psi_t^{{\rm norm}}(s) \rangle 
\nonumber \\
&=
\frac{1}{Z_{t}(s)} \,
\sum_{K=0}^{\infty}  \,
\sum_{\mu_{1} \cdots \mu_{K}=1}^{\NJ} \,
\int_{0 \leq t_1 \cdots t_{K} \leq t} \,
e^{{-s A(\omega_{t_0})}} 
\langle - | e^{-s A(\omega_+)}
\V \left( \omega_{+} \right) \V \left( \omega_{t_0} \right) 
| C_{0} \rangle \otimes | \omega_{t_0} \rangle 
\nonumber \\
&=
\frac{1}{Z_{t}(s)} \,
\sum_{K_0=0}^{\infty} \,
\sum_{\mu_{1} \cdots \mu_{K_0}=1}^{\NJ} \,
\int_{0 \leq t_1 \cdots t_{K_0} \leq t_0} \,
e^{{-s A(\omega_{t_0})}} 
\langle - | 
e^{(t-t_0) \L(s)} \V \left( \omega_{t_0} \right) 
| C_{0} \rangle \otimes | \omega_{t_0} \rangle ,
\nonumber
\\
&=
\frac{e^{-t_0 \theta(s)}}{\langle l_s | C_0 \rangle}
\sum_{K_0=0}^{\infty}  \,
\sum_{\mu_{1} \cdots \mu_{K_0}=1}^{\NJ} \,
\int_{0 \leq t_1 \cdots t_{K_0} \leq t_0} \,
e^{{-s A(\omega_{t_0})}} 
\langle l_s | \V \left( \omega_{t_0} \right) 
| C_{0} \rangle \otimes | \omega_{t_0} \rangle .
\label{recn}
\end{align}
Between the second and third lines we have integrated over all the possible trajectories $\omega_+$ after $t_0$ and have used \er{NS2}.  We have labelled the sum over jumps with the index $K_0$ to indicate that this sum, and the time integrals, correspond to the sub-trajectories up to $t_0$ only.  To get the final line we use that $t$ is large, with $t_0 \ll t$, and \era{Ns}{LD}. 
We proceed in a similar way with the microcanonical cMPS \er{amcn} to get,
\begin{align}
| \psi_{t_0}^A \rangle 
&= 
\langle - | \otimes \langle -_{{\rm out}_+} | \Psi_t^{A,{\rm norm}} \rangle 
\nonumber \\
&=
\frac{1}{P_{t}(A)} \,
\sum_{K=0}^{\infty}  \,
\sum_{\mu_{1} \cdots \mu_{K}=1}^{\NJ}
\int_{0 \leq t_1 \cdots t_K \leq t}
\delta[ A - A(\omega_{t_0}) - A(\omega_+)] \,
\langle - | 
\V \left( \omega_{+} \right) \V \left( \omega_{t_0} \right) 
| C_{0} \rangle \otimes | \omega_{t_0} \rangle 
\nonumber \\
&=
\frac{1}{P_{t}(A)} \,
\sum_{K_0=0}^{\infty} \,
\sum_{\mu_{1} \cdots \mu_{K_0}=1}^{\NJ}
\int_{0 \leq t_1 \cdots t_{K_0} \leq t_0}
\sum_C
P_{t-t_0}[A - A(\omega_{t_0}) | C] \,
\langle C | 
 \V \left( \omega_{t_0} \right) 
| C_{0} \rangle \otimes | \omega_{t_0} \rangle 
\nonumber \\
&=
\frac{e^{t \varphi\left(\frac{A}{t}\right)}}{\langle l_s | C_0 \rangle}
\sum_{K_0=0}^{\infty} \,
\sum_{\mu_{1} \cdots \mu_{K_0}=1}^{\NJ}
\int_{0 \leq t_1 \cdots t_{K_0} \leq t_0}
e^{-(t-t_0) \varphi\left(\frac{A - A(\omega_{t_0})}{t-t_0}\right)}
\langle l_s | \V \left( \omega_{t_0} \right) 
| C_{0} \rangle \otimes | \omega_{t_0} \rangle .
\label{remcn}
\end{align}
As before, we get the third line after integrating over $\omega_+$.  Here $P_{t-t_0}[A - A(\omega_{t_0}) | C]$ indicates the probability of observing $A - A(\omega_{t_0})$ after time $t-t_0$ starting from configuration $C$.  The meaning should be clear: for the sub-trajectory $\omega_{t_0}$ the observable up to time $t_0$ is $A(\omega_{t_0})$, so the trajectory after $t_0$ must have $A - A(\omega_{t_0})$, since trajectories are conditioned on having a total of $A$.  The final line is obtained by again considering the large $t$ limit and \er{LD}. 
If we also assume that $A$ is large, we can approximate the rate function $\varphi\left[A - A(\omega_{t_0})/t-t_0\right]$ in the integrand.  Since $t_0 \ll t$ this function will be only significant for $A(\omega_{t_0}) \ll A$.  We can then expand:
\begin{equation}
- \left( t-t_0 \right)
\varphi\left(\frac{A - A(\omega_{t_0})}{t-t_0}\right)
=
- \left( t-t_0 \right)
\varphi\left(\frac{A}{t}\right) 
+ A(\omega_{t_0}) 
\varphi'\left(\frac{A}{t}\right)
- t_0 \frac{A}{t} 
\varphi'\left(\frac{A}{t}\right)
+ \cdots 
\end{equation}
The reduced microcanonical state \er{remcn} becomes,
\begin{align}
| \psi_{t_0}^A \rangle 
&=
\frac{e^{t_0 \left[ 
\varphi\left(\frac{A}{t}\right) - \frac{A}{t} 
\varphi'\left(\frac{A}{t}\right)
\right]}}{\langle l_s | C_0 \rangle}
\sum_{K_0=0}^{\infty} 
\sum_{\mu_{1} \cdots \mu_{K_0}=1}^{\NJ}
\int_{0 \leq t_1 \cdots t_{K_0} \leq t_0}
e^{A(\omega_{t_0})  \varphi'\left(\frac{A}{t}\right)}
\langle l_s | \V \left( \omega_{t_0} \right) 
| C_{0} \rangle \otimes | \omega_{t_0} \rangle ,
\label{remcn2}
\end{align}
\end{widetext}
If we now take $A \to \infty$ and $t \to \infty$ such that $A/t = -\theta'(s)$, cf.\ \er{as}, the two reduced cMPSs, \era{remcn2}{recn}, become identical.   This means that the distribution of all finite-time sub-trajectories from an ensemble conditioned on $A$ on long times is the same as that from ensemble conditioned via $s$, as long as $A$ and $s$ are related by \er{as}.  Microcanonical and canonical trajectory ensembles are therefore equivalent \cite{Chetrite2013}.

\section{Integral fluctuation theorems from gauge transformations}

We finish by discussing how integral fluctuation relations \cite{Seifert2012} can be proven through the gauge symmetry of the cMPSs encoding the dynamics.  A general discussion on the connection between gauge invariance in stochastic dynamics and fluctuation theorems was given in Ref.\ \cite{Chetrite2011}.  

For concreteness we focus on the steady state fluctuation theorem for the entropy production into the environment of Lebowitz and Spohn \cite{Lebowitz1999}.  We consider dynamics \ers{ME2}{L} where $\{ \W_{C \to C'} \}$ and $\R$ do not depend on time, and there is a stationary distribution $\L | P_{\rm st} \rangle = 0$ (which in general will not be an equilibrium one as detailed balance need not be obeyed).  Whenever a jump $C \to C'$ occurs in a trajectory the entropy produced in the environment is \cite{Seifert2012}
\begin{equation}
\Delta S_{C \to C'} = k_{\rm B} \log \left( 
\frac{W_{C \to C'}}{W_{C' \to C}} \right)
\label{Scc} ,
\end{equation}
where $k_{\rm B}$ is Boltzmann's constant and $W_{C' \to C}$ is the rate for inverse transition (and we assume that if $W_{C \to C'} > 0$ then $W_{C' \to C} > 0$ as well for all pairs $C,C'$).  The total entropy produced by a trajectory is therefore the sum of these contributions over all the jumps.  From \er{AA} we can then write an integrated environment entropy production output operator as 
\begin{equation}
\Delta \S = \sum_{\{ C \to C' \}} \int_0^t dt' \, \Delta S_{C \to C'}  \, a_{C \to C'}^\dagger(t') \, a_{C \to C'}(t') ,
\label{S}
\end{equation}
where we have made explicit that the sum over all possible jumps $\mu$ corresponds to the sum over all possible configuration transitions $C \to C'$.  Consider the cMPS
of an ensemble of trajectories conditioned on the average of $\Delta \S$,
\begin{equation}
| \Psi_t(\lambda) \rangle = e^{- \lambda \Delta \S} | \Psi_t \rangle ,
\label{psila}
\end{equation}
where $\lambda$ is the corresponding conjugate field. 
This biased ensemble is generated by the tilted operator \cite{Lebowitz1999} [cf.\ \er{MPS} and \er{Ls}]
\begin{equation}
\L(\lambda) = \sum_{\{ C \to C' \}} 
\W_{C \to C'}  
\left( 
\frac{W_{C \to C'}}{W_{C' \to C}} \right)^{-\lambda}
 - \R ,
 \label{Llambda}
\end{equation}
where we have set $k_{\rm B}=1$. 

Since, cf.\ \er{Ns},
\begin{equation}
\langle -_{\rm tot} | \Psi_t(\lambda) \rangle = 
\left< e^{- \lambda \Delta \S} \right> ,
\label{avla}
\end{equation}
the fluctuation theorem of Ref.\ \cite{Lebowitz1999},
\begin{equation}
\left< e^{-\Delta \S} \right> = 1 ,
\label{DS}
\end{equation}
corresponds to the statement that $| \Psi_t(1) \rangle$ is equivalent to the cMPS of an actual stochastic process (without the need for normalisation of the conditioned cMPS as in Sec.\ IV).   This is easily proven via a gauge transformation \er{G} of the cMPS $| \Psi_t(1) \rangle$.  Requiring stochasticity, cf.\ \er{ts}, results in the gauge fixing condition, cf.\ \er{eqG}, in the case where the transformation is time-independent and $f=0$,
\begin{equation}
\langle - | \G \, \L(1) = 0 .
\label{gla}
\end{equation}
This condition is satisfied for
\begin{equation}
\G | - \rangle = | P_{\rm st} \rangle ,
\end{equation}
given that from \er{Llambda} we have that $\L(1) = \L(0)^\dagger$.  That is, the conditioned ensemble encoded in $| \Psi_t(1) \rangle$ is equivalent, via a gauge transformation, to one generated by a proper stochastic dynamics.  This then implies the existence of an associated integral fluctuation relation, which in this particular case is \er{DS}.  Other fluctuation relations are obtained in a similar manner \cite{Chetrite2011}: The trajectory ensemble is reweighted with the exponential of a trajectory observable (such as entropy production, work, etc.).  This defines the corresponding tilted cMPS, cf.~\er{psila}, whose norm is the average of the exponential of the observable, cf.~\er{avla}.  If the corresponding conditioned process can be mapped to a stochastic one via a gauge transformation then we have an integral fluctuation theorem, cf.~\era{DS}{gla}.

\section{Conclusions}

We have introduced a formalism based on matrix product states to catalog ensembles of trajectories in classical stochastic systems.  This approach  is a classical version of the known connection between continuous MPSs and open quantum dynamics \cite{Verstraete2010,Haegeman2013,Osborne2010}. It allows to describe in a compact way conditioned trajectory ensembles and demonstrate ensemble equivalences, in analogy with what can be done in the quantum case \cite{Lesanovsky2013,Kiukas2015}. 
The key property of cMPSs is that of gauge invariance from which the equivalences follow.  Other dynamical properties can be proved via gauge transformations, as for example certain fluctuation relations \cite{Chetrite2011}. 

We have focused on systems evolving as continuous time Markov chains.  For this kind of dynamics the corresponding MPS that encodes the set of trajectories is a continuous MPS given that jump events can occur at any time.  One can describe also discrete Markov chain dynamics with standard MPSs where the auxiliary space is a discrete lattice rather than the real line.  Other generalisations are possible.  For example, here we have considered trajectories which terminate at some fixed maximum time, but one could  consider instead the set of trajectories which terminate after a fixed number of jumps and where their total time extent is variable \cite{Bolhuis2008,Budini2014}.   In analogy with the quantum case \cite{Kiukas2015}, we expect such a variable time ensemble to have a discrete MPS description. 

The cMPS vectors encode all the information about the dynamics.  The various levels of large-deviations \cite{Touchette2009,Baiesi2009b,Bertini2012} can therefore be recovered by contraction of the output space: for example, the contraction $| \omega_{t} \rangle \to | A(\omega_t) \rangle$ from an output state that encodes the whole trajectory $\omega_{t}$ to one that only retains the value of a trajectory observable, cf.\ \er{A}, would
produce a reduced system-output vector corresponding to a level-1 (or level-1.5) description \cite{Touchette2009}.  Similarly, a contraction $| \omega_{t} \rangle \to | \{ n_\mu \} \rangle$ where the total number of jumps of each kind $\mu$ is retained would result in a system-output vector corresponding to a level-3 description.   In this way one can expect to prove various large-deviation variational relations  \cite{Touchette2009,Baiesi2009b,Bertini2012} within the MPS approach presented here.

\bigskip

\begin{acknowledgements}
This work was supported in part by EPSRC Grant No.\ EP/K01773X/1.

\end{acknowledgements}

\bibliography{gauge}

\end{document}